\documentstyle[aps,prb,epsf,amsfonts]{revtex}

\begin{document}
\draft
\preprint{{\bf ETH-TH/98-06}}

\wideabs{
\title{Numerical study of vortex matter using the Bose model:\\
       First-order melting and entanglement}
\author{Henrik Nordborg$^{a,b}$ and Gianni Blatter$^{a}$}
\address{$^{a}$Theoretische Physik, ETH-H\"onggerberg, CH-8093
         Z\"urich, Switzerland \\
         $^{b}$Materials Science Division, Argonne National
         Laboratory, Illinois 60469}
\date{\today}

\maketitle

\tighten

\begin{abstract}
We present an extensive numerical study of vortex matter using the
mapping to 2D bosons and Path-Integral Monte Carlo simulations. We
find a \emph{first-order} vortex lattice melting transition into an
\emph{entangled} vortex liquid. The jumps in entropy and density are
consistent with experimental results on YBa$_2$Cu$_3$O$_{7-\delta}$.
The liquid is denser than the lattice and has a correlation length
$l_z \approx 1.7\varepsilon a_0$ in the direction parallel to the
field. In the language of bosons we find a sharp quantum phase
transition from a Wigner crystal to a superfluid, even in the case of
logarithmic interaction. We also measure the excitation spectrum of
the Bose system and find the roton minimum to be insensitive to the
range of the interaction.
\end{abstract}

}

\narrowtext

\section{Introduction}

The Abrikosov mean-field theory of the vortex system predicts a
peculiar continuous freezing transition, involving two symmetries, at
the upper critical field $H_{c2}$\cite{Abrikosov:JETP57}. The gauge
symmetry is broken in the direction parallel to the field by the
appearance of a superconducting order parameter $\Psi({\bf x})$. The
presence of the vortex lattice, on the other hand, leads to a
periodic modulation of the amplitude of the order parameter, breaking
the translational invariance of the system. This behavior contradicts
a general argument by Landau, stating that breaking of continuous
translation symmetry should be associated with a first order
transition\cite{Landau:JETP37b}. In order to resolve the paradox, we
have to go beyond mean-field theory and take thermal fluctuations
into account: The fluctuations melt the vortex lattice in a
first-order transition at temperatures below the mean-field
transition in agreement with the Landau
argument\cite{Brezin-etal:PRB85}. The size of the region where
thermal fluctuations are important is determined by the Ginzburg
criterion: In conventional superconductors this region is small and
mean-field theory offers a good description of the relevant physics.
In the high-$T_c$ materials, on the other hand, thermal fluctuations
drastically change the mean-field phase diagram. Understanding the
effect of thermal fluctuations is therefore vital for the
understanding of the vortex state in high-$T_c$ superconductors.

There exist a number of fundamental questions associated with the
melting of the vortex lattice. To begin with, the existence of two
symmetries naturally leads to the question of whether these have to
be broken simultaneously in one transition\cite{Blatter-etal:RMP94}.
In terms of vortices, the gauge symmetry along the field is destroyed
by the alignment of the vortices with the applied magnetic field,
whereas the translational symmetry is broken by the ordering of the
lines in the direction transverse to the field. A melting scenario
involving a disentangled liquid, which has been the topic of an
intense scientific discussion over the past
years\cite{Feigelman-etal:PRB93,Chen-Teitel:PRL96comment}, would
break these symmetries in two separate transitions. A further topic
deals with the properties of the melted phase: Is the resulting
\emph{vortex liquid} thermodynamically identical to the normal state
or do further phase transitions exist at higher temperatures.

Although the importance of thermal fluctuations in the vortex system
was recognized early%
\cite{Eilenberger:PR67,Labusch:PSS69,Brezin-etal:PRB85}, most
progress has been made only in recent years. One important reason for
this is that it has now become possible to observe the 
vortex lattice melting transition experimentally: After the first
indirect observations based on resistivity measurements%
\cite{Safar-etal:PRL92,Charalambous-etal:PRB92}, more recent
experiments have observed the first-order transition directly by
measuring 
a jump in the magnetization or the latent heat%
\cite{Zeldov-etal:NAT95,Welp-etal:PRL96,%
Schilling-etal:NAT96,Roulin-etal:SCI96}. 
The very large latent heat which was measured in many experiments was
puzzling at first, but the problem has now been resolved
satisfactorily%
\cite{Hu-MacDonald:PRB97,Dodgson-etal:PRL98}. 

On the theoretical side, there still does not exist a reliable theory
for the vortex lattice melting transition. Early work using the
renormalization group\cite{Brezin-etal:PRB85} or density functional
theory\cite{Sengupta-etal:PRL91} have indicated a first-order
transition. Elastic theory has been combined with the Lindemann
criterion to produce a melting line in good agreement with
experimental observations%
\cite{Houghton-etal:PRB89,Nordborg-etal:PRB96}. Even
though the Lindemann criterion does not give explicit information on
the character of the melting line, it can be argued that the fact
that it works so well gives support for a first-order transition. The
question regarding the possible existence of a disentangled liquid
has been addressed by analyzing the energy and stability of entangled
configurations both in the liquid and in the
lattice\cite{Schoenenberger-etal:PRL95}, with the conclusion that no
disentangled liquid should exist.

The absence of an analytical description of the vortex lattice
melting transition has led to a large interest in numerical
simulations. A popular approach is to use the frustrated
XY-model%
\cite{Li-Teitel:PRL91,Hetzel-Sudbo-Huse:PRL92,Li-Teitel:PRB93,%
Chen-Teitel:PRB97,Hu-etal:PRL97,Koshelev:PRB97,%
Nguyen-Sudbo:PRB98a,Nguyen-Sudbo:PRB98b} or the closely
related Lattice London model%
\cite{Chen-Teitel:PRL95,Carneiro:PRL95,Sudbo-etal:PRL96,%
Hagenaars-etal:PRB97}. Unfortunately, many of the simulations have
suffered from highly non-trivial finite-size effects and have
indicated two transitions instead of one at low filling factors,
whereas a first-order transition was seen at large filling
factors\cite{Hetzel-Sudbo-Huse:PRL92}. The problems have been
overcome recently\cite{Hu-etal:PRL97}, and a picture with a single
first-order transition is emerging. An overview of these problems can
be found in Ref.~\onlinecite{Koshelev:PRB97}. Simulations based on
the Lowest Landau Level (LLL) approximation have shown a first-order
vortex lattice melting transition\cite{Sasik-Stroud:PRL95} and
obtained entropy jumps in agreement with
experiments\cite{Hu-MacDonald:PRB97}.

In this work we adopt a different approach from those described
above. It was suggested by Nelson that the vortex system is
equivalent to a system of interacting bosons in two
dimensions\cite{Nelson:PRL88}. This idea allows for using standard
many-body techniques in describing the vortex system and has become a
popular approach for analytical work on
vortices%
\cite{Nelson-Vinokur:PRB93,%
Hatano-Nelson:PRL96,Tauber-Nelson:PhysRep97}. 
The mapping to bosons, which we shall refer to as the \emph{Bose
model} in the following, has not been widely used for numerical work,
however. This is perhaps surprising, since there exist numerical
algorithms for simulating interacting bosons exactly. The main
advantage of the Bose
model lies in its generality: By analyzing this model, which contains
very few adjustable parameters, we are able to obtain results for
vortices, bosons, and interacting elastic lines, such as polymers, in
general. Our approach also allows us to answer two questions which
cannot be considered with the methods above. The first is obvious:
How well does the Bose model describe the real vortex lattice? This
has not been studied before, as it is difficult to make any
quantitative predictions for strongly interacting bosons
analytically. The second question is of a more fundamental nature: It
has been known for a while that the 2D Bose Coulomb Liquid does not
have a Bose-Einstein condensate even at 
$T = 0$%
\cite{Campbell:78,Pitaevskii-Stringari:JLTP91,Magro-Ceperley:PRL94}.
Consequently, it has  been questioned whether this system necessarily
has to be crystalline or superfluid in its ground state, or whether
some other phase might exist \cite{Feigelman-etal:PRB93}. Here we show
that the system has a direct transition from a crystal to a
superfluid, and that a superfluid without a Bose-Einstein condensate
can indeed exist.

Some early results of this work have been published previously%
\cite{Nordborg-Blatter:PRL97}. 
Here we expand on our previous work, adding new results for
melting in a compressible system using an isobaric Monte Carlo 
method. These results allow us to use the Clausius-Clapeyron
relation as an consistency check on the simulation, since the 
change in density and the change in entropy can be measured
independently at the transition. New results for
the energy of corresponding Bose system are presented,
and we discuss all results both in the language of 
bosons and vortices, further emphasizing similarities 
and differences between the two systems.
Finally, specific details on the simulation are provided.

The outline of the paper is as follows: In Sec.~\ref{secmodel} we
describe the model and discuss its applicability to the vortex
problem and the approximations made. We then turn to a qualitative
description of the bosonic phase-diagram in Sec.~\ref{secbosons}, and
relate it to the phase diagram for vortices. The algorithm and the
simulations are discussed in Sec.~\ref{secalgo}. Numerical results
for the 2D Bose system are given in Sec.~\ref{sec2dbcl} and the
numerical results for the vortex lattice melting transition are
presented in Sec.~\ref{secmelting}. In Sec.~\ref{secliquid} we look
at some properties of the vortex liquid and our conclusions are
summarized in Sec.~\ref{secconclude}. Three appendices provide
technical details on various aspects of the simulation. 

\section{Description of the model}
\label{secmodel}

\subsection{Boson-vortex mapping}

We consider a system of $N$ bosons in two dimensions interacting
through the potential $V(r) = g^2 {\rm K}_0(r/\lambda)$, where $g^2$
is the energy scale of the interaction, ${\rm K}_0(x)$ is a modified
Bessel function of the second kind, and $\lambda$ measures the range
of the interaction. In the Feynman path-integral
picture\cite{Feynman:72}, the system is described by interacting
``elastic'' world-lines with the \emph{imaginary-time} action 
\begin{equation}
\frac S \hbar =  \frac 1 \hbar \int_0^{\hbar / T} d\tau \left\{
\sum_i \frac m 2 \left(\frac{d{\bf R}_i}{d \tau} \right)^2 
+ \sum_{i < j} g^2{\rm K}_0 \left( \frac{R_{ij}} {\lambda}\right)
\right\}.
\label{boseaction}
\end{equation}
Here, the two-dimensional vectors ${\bf R}_i(\tau)$ represent the
positions of the bosons as a function of the imaginary time $\tau$
and $T$ is the temperature of the system. The boundary condition for
bosons is ${\bf R}_i(0) = {\bf R}_j(\beta)$, i.e., every line ends
either on itself or on some other line. The partition function is the
average of all possible line configurations subject to this boundary
condition and weighted by the action (\ref{boseaction}):
\begin{equation}
{\cal Z}(T, A, N) \equiv {\rm e}^{-F(T,A,N)/T} = \frac 1 {N!} \int
\prod_{i=1}^N {\cal D}{\bf R}_i \, {\rm e}^{-S/\hbar},
\label{partfunc}
\end{equation}
where $F(T,A,N)$ denotes the Helmholtz free energy and $A$ is the
area of the system (we are considering two dimensions). It is often
useful to consider a system of bosons at fixed chemical potential
$\mu$ rather than at fixed number of particles. Switching ensemble is
easily done according to 
\begin{equation} 
{\cal Z}(T,A,\mu) \equiv {\rm e}^{-\Omega(T,A,\mu)/T} = \sum_{N =
0}^\infty {\cal Z}(T,A,N){\rm e}^{\mu N/T},
\label{GrandCanDef}
\end{equation}
in which case $\Omega(T,A,\mu)$ denotes the grand-canonical
potential.

It was pointed out by Nelson, that the action in
Eq.~(\ref{boseaction}) can also be interpreted as the London
free-energy for a system of interacting vortices in type-II
superconductors\cite{Nelson:PRL88,Nelson-Seung:PRB89}. The starting
point is the London free-energy functional 
\begin{equation}
\frac {\cal F} T = \frac 1 T \int_0^{L_z} \!\! dz \left\{ \sum_i
\frac {\varepsilon_l} {2} \left( \frac{d{\bf R}_i}{d z} \right)^2 + 
\sum_{i < j} 2 \varepsilon_0 {\rm K}_0 \left( \frac{R_{ij}}
{\lambda}\right)
\right\},
\label{GLfree}
\end{equation}
where $\Phi_0$ is the flux quantum, $\varepsilon_0 = \Phi_0^2 / (4\pi
\lambda)^2$ is the vortex line energy, $L_z$ is the thickness of the
sample, and $\varepsilon_l$ is the elasticity of a vortex line. The
latter depends strongly on the anisotropy and we have the estimate
$\varepsilon_l \approx \varepsilon^2 \varepsilon_0 \ln( a_0/ 2
\sqrt{\pi}\xi)$, where $\varepsilon^2 < 1$ is the anisotropy
parameter, $\xi$ is the coherence length and $a_0$ is the lattice
spacing. A detailed discussion of the line energy is given in
Sec.~\ref{secvalidity}. We will refer to Eq.\ (\ref{GLfree})
as the Bose model for the vortex system in the following. 

Comparing Eqs.\ (\ref{boseaction}) and (\ref{GLfree}), and requiring
${\cal F} / T = S/\hbar$, we obtain the following formal mapping
between 2D bosons and vortices:
\begin{equation}
\begin{array}{ccc} 
\tau & \leftrightarrow & z  \\ 
\hbar & \leftrightarrow & T \\
g^2/\hbar & \leftrightarrow & 2\varepsilon_0  \\
\hbar / T_B & \leftrightarrow & L_z \\
m & \leftrightarrow & \varepsilon_l 
\end{array}
\end{equation}
The temperature of the Bose system is denoted by $T_B$, in order to
distinguish it from the (physically different) temperature $T$ of
the vortex system. Some conclusions can immediately be drawn from
this mapping: The temperature of the vortex system corresponds to the
Planck constant of bosons, implying that \emph{thermal fluctuations}
of vortices are equivalent to \emph{quantum fluctuations} of bosons.
Quantum fluctuations are more relevant for bosons with small mass.
Since the mass of the bosons corresponds to the elasticity of the
vortices, it follows that thermal fluctuations are more important for
vortices with small elasticity, i.e., for anisotropic materials.
Finally, the vortex system in a bulk superconductor, with $L_z
\rightarrow \infty$, corresponds to the ground state, $T_B = 0$, of
the bosons. 

As in the case with bosons, it is often useful to consider a system
where the number of vortices is allowed to fluctuate. The number of
vortices determines the magnetic field and we therefore have to
consider the Gibbs energy 
\begin{equation}
{\cal G} = {\cal F} - \int d^3r \, \frac{HB}{4\pi}.
\end{equation} 
This quantity should be compared with $S/\hbar - N \mu / T_B$ for the
bosons, and we obtain the formal equivalence
\begin{equation}
\frac \mu \hbar \leftrightarrow \left( \frac{\Phi_0 H}{4\pi} -
\varepsilon_0
\ln \kappa \right).
\label{chempot}
\end{equation}
The chemical potential $\mu$ is negative for $H < H_{c1} = \Phi_0 \ln
\kappa / 4\pi\lambda^2$, in agreement with the fact that no vortices
enter the sample for fields below the lower critical field.

It will be convenient to introduce dimensionless parameters in the
following analysis, which is done by choosing a length and an energy
scale. As the length scale we take the lattice constant $a_0$ of a
triangular lattice, i.e., the density is $\rho = 2/a_0^2\sqrt{3}$.
The natural choice for the energy scale is $g^2$ for bosons and
$\varepsilon_0 a_0$ for the vortex system. If we consider a fixed
number of particles, where the chemical potential is not needed, the
thermodynamic properties of the system are completely determined by
the \emph{three dimensionless parameters}
\begin{mathletters}
\label{parammap}
\begin{eqnarray}
\Lambda & \equiv & \frac{\hbar}{a_0g\sqrt{m}} =
\frac{T}{a_0\sqrt{2\varepsilon_l\varepsilon_0}}, \\
\beta & \equiv & \frac{g^2}{T_B} = \frac{2\varepsilon_0 L_z}{T}, \\
\tilde{\lambda} & \equiv & \lambda / a_0,
\end{eqnarray}
\end{mathletters}
with the dimensionless action
\begin{equation}
S = \int_0^{\beta} \!\! d\tau \left\{ \sum_i
\frac {1} {2\Lambda^2} \left( \frac{d{\bf R}_i}{d \tau} \right)^2 + 
\sum_{i < j} {\rm K}_0 \left( R_{ij} / \tilde{\lambda} \right)
\right\}.
\end{equation}
The parameter $\Lambda$ is the de Boer parameter and measures the
size of fluctuations; \emph{quantum} for bosons and
\emph{thermal} for vortices. This is the most important parameter in
the present study. $\beta$ is the dimensionless inverse temperature
and we will only be considering the limit $\beta \rightarrow \infty$
here, corresponding to the ground state for bosons or vortices in a
bulk superconductor. Finally, it is often a good approximation to use
$\lambda = \infty$, which further reduces the number of parameters.
In the latter case, we are studying the two-dimensional Bose Coulomb
liquid (2DBCL) with logarithmic interaction. Note that simply taking
the limit $\lambda \rightarrow \infty$ in the action
(\ref{boseaction}) leads to a diverging energy and the interaction
with a background charge therefore has to be subtracted. The density 
is then fixed by the requirement of 
charge neutrality and the system is therefore 
\emph{incompressible}. In order to have a finite
compressibility, which is needed for a density (or magnetization)
jump at a first-order phase transition, we have to use a finite value
of $\lambda$. Since the vortices in the XY-model interact
logarithmically, it follows that this model will never have a
magnetization jump, in contrast to recent
claims\cite{Ryu-etal:PRL97}.

\subsection{Validity of the model}
\label{secvalidity}

Before discussing the phase diagram of the model described above, we
want to discuss its validity as a model for vortices in a type II
superconductor. It is natural to divide this discussion into three
parts, considering boundary conditions, linearization, and retarded
interactions separately.

The Bose model differs from a real vortex system in the choice of
boundary conditions: The natural boundary conditions for vortices are
given by 
\begin{equation}
\frac{d{\bf R}_i}{dz}(0) = \frac{d{\bf R}_i}{dz}(L_z) = 0,
\end{equation}
since the screening currents, which encircle the vortex, cannot leave
the sample. Since one cannot hope to simulate a sample with realistic
dimensions we make use of periodic boundary conditions in all
directions, as this minimizes the effect of the boundaries 
by ensuring $\nabla\cdot{\bf B} = 0$ everywhere. The
expectation then is that the extrapolation of our numerical results
to infinite system size provides an accurate description of the 
transition. The bosonic boundary conditions in the longitudinal 
direction, ${\bf R}_i(\beta) = {\bf R}_j(0)$, correspond to 
periodic boundary conditions for the magnetic field in the 
$z$-direction and have been used in most numerical simulations of 
vortices in bulk materials so far.

Another question is related to the linearization leading to the
elastic term in Eq.~(\ref{GLfree}), i.e.,
\begin{equation}
\frac{dl}{dz} = \sqrt{1 + \left( \frac{d{\bf R}}{dz} \right)^2}
\approx 1 +
\frac 1 2 \left( \frac{d{\bf R}}{dz} \right)^2
\end{equation}
for an isotropic superconductor. This expansion is only valid as long
as the vortex lines do not deviate too much from the $z$-axis. It is
very accurate, however: For a vortex at a $30^\circ$ angle with the
$z$-axis, the error is roughly $1\%$. Close to the melting transition
we have $\left( d{\bf R}/dz \right)^2 \approx c_L^2$, showing that
the Taylor expansion is valid. Scaling theory proves this result to
be true also in an anisotropic
superconductor%
\cite{Blatter-Geshkenbein-Larkin:PRL92,Nelson-Vinokur:PRB93}.

The use of the $z$-coordinate as a parameter for the vortex lines
prohibits the formation of vortex loops in the $ab$-planes. 
Since these loops have been argued to be important for vortex lattice
melting\cite{Sudbo-etal:PRL96,Nguyen-Sudbo:PRB98a}, we would like to 
remark on this point here. The free energy for a planar vortex loop 
of length $L$ is given by
\begin{equation}
F_L = \left( \frac{L}{\xi} \right) \left[ \varepsilon \varepsilon_0
\xi - T \ln 3\right],
\end{equation}
if we use a simple random walk argument for estimating the entropy.
Loops will begin to proliferate when the free energy becomes
negative, i.e., for $T > \varepsilon \varepsilon_0 \xi / \ln 3$.
This, however, is essentially the Ginzburg criterion, showing that
planar vortex loops are important in the fluctuation regime close to
the upper critical field $H_{c2}$. Due to the smallness of the
Lindemann number for vortex lattice melting, the melting line is
usually located well outside the fluctuation
regime\cite{Blatter-etal:RMP94,Nordborg-etal:PRB96}.

A more important approximation is the way the Bose model, i.e.,
Bosons with Yukawa interaction, neglects the ``retarded'' interaction
of the real vortex lattice by the
substitution\cite{Blatter-etal:RMP94} 
\begin{equation}
\int dz' \frac{{\rm
e}^{-\sqrt{R_{ij}(z,z')^2+(z-z')^2}/\lambda}}{\sqrt{R_{ij}(z,z)^2
+(z-z')^2}} \Rightarrow {\rm K}_0\left[ R_{ij}(z,z) /
\lambda\right].
\label{retarded}
\end{equation}
We have used the notation $R_{ij}(z,z') = |{\bf R}_i(z) - {\bf
R}_j(z')|$. Again, this approximation is valid as long as the lines
are straight, at least over distances $\lambda$ in the $z$-direction.
However, Eq.~(\ref{retarded}) replaces a non-local interaction in the
$z$-direction with a local one. The consequences of this can be
understood in terms of the elastic properties of the vortex lattice:
Since it is the non-local character of the interaction which produces
dispersion, it follows that the substitution (\ref{retarded}) will
remove any $k_z$-dependence in the elastic moduli. The shear modulus
$c_{66}$ is not dispersive and remains unchanged. The compression
modulus $c_{11}$, which is not important for vortex lattice melting,
is changed in a trivial way. The tilt modulus, however, deserves a
more detailed treatment. It is easy to see that the tilt modulus
corresponding to Eq.~(\ref{GLfree}) is given by \begin{equation}
c_{44} = \frac{2}{\sqrt{3}} \frac{\varepsilon_l}{a_0^2},
\label{bosetilt}
\end{equation}
because only the self-energy terms contribute to the tilt energy. The
real tilt modulus of the vortex lattice is rather complicated,
especially in the case of layered materials%
\cite{Sudbo-Brandt:PRL91,Glazman-Koshelev:PRB91,Nordborg-etal:PRB96}.
The single vortex contribution is given by 
\begin{equation}
c_{44}^0(K,k_z) \approx \frac{\varepsilon^2 \varepsilon_0}{2a_0^2} \ln
\left\{
\frac{ \left(\kappa/\varepsilon\right)^2 }{1  + \lambda^2
K_{\scriptscriptstyle BZ}^2 /
\varepsilon^2 + \lambda^2 k_z^2 } \right\}
\label{linetilt}
\end{equation}
and has the same form as Eq.~(\ref{bosetilt}) apart from a
logarithmically weak $k_z$-dispersion. In addition, the real vortex
system has a bulk contribution to the tilt, emanating from the
interaction between vortices, which is given by
\begin{equation}
c_{44}(K,k_z) = \frac{B^2}{4\pi}\frac{1}{1 + \lambda^2 K^2 /
\varepsilon^2 +
\lambda^2 k_z^2}.
\label{tiltmod}
\end{equation}
This term is absent in the Bose model, since the interaction energy
only depends on the distance between lines and not on their angle with
the applied field. For thermal fluctuations, the relevant modes have
wave vectors $K \approx K_{\scriptscriptstyle BZ} \approx \sqrt{4\pi}
/ a_0$ and
$k_z \approx K / 2 \varepsilon$. Whence, the single-vortex part of the
tilt
modulus gives the dominant contribution to the stiffness of the
lattice and we expect the Bose model to be valid for describing 
thermal fluctuations as long as we use Eqs.~(\ref{bosetilt}) and
(\ref{linetilt}) to determine $\varepsilon_l$, i.e., by choosing
$\varepsilon_l =
\varepsilon^2\varepsilon_0 \ln( a_0 / 2\sqrt{\pi} \xi )$.

The above discussion made use of the elastic moduli of the vortex
lattice and is not directly applicable to the liquid. We do not know
enough about the properties of the vortex liquid to be able to make an
analytical comparison between the Bose model and the London model in
this case. A major difference is that the vortex lines in the Bose
model always repel each other, whereas the interaction between real
vortices is proportional to the cosine of the angle between the lines
and can become attractive. The difference is not important
in the lattice, since the vortices are almost parallel anyway. In the
liquid the vortices are not so well aligned and the difference will
be larger. We argued above that the linearization of the self energy
is very accurate, since the error is a fourth-order correction in
$dR_i/dz$. Similarly, the error done by dropping the cosine is a
second-order correction in the same quantity. The Bose model
therefore still contains the main part of the interaction between the
vortices and we expect our results to be in rough quantitative
agreement with those of the real system. This is further supported by
the fact that our model correctly captures the properties of the
vortex lattice melting transition, as will be shown below. 

A very interesting consequence of the direction-dependent interaction
for real vortices is the existence of an attractive van der Waals
interaction even for straight vortices. This attractive interaction
has important consequences for the low-field phase diagram of
anisotropic superconductors, below the lower melting
line\cite{Blatter-Geshkenbein:PRL96}. For the magnetic inductions
considered here, however, with $a_0 \lesssim \lambda$, the van der
Waals interaction can safely be ignored.

\section{The phase diagram for bosons with long-range interaction}
\label{secbosons}

\begin{figure}
\centerline{\epsfxsize= 8.5cm\epsfbox{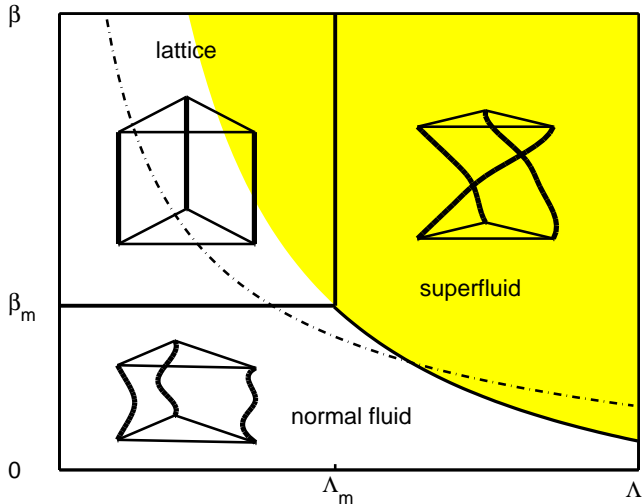}}
\caption{Schematic phase diagram of the 2DBCL in the parameters
$\Lambda$ and $\beta$. The solid lines indicate real phase
transitions and quantum effects are important in the shaded region.
For vortices in high-$T_c$ superconductors we have $\Lambda^2 = T^2/2
\varepsilon_l \varepsilon_0 a_0^2$ and $\beta = 2\varepsilon_0 L_z /
T$. The dashed line is obtained by keeping $B$ ($a_0$) fixed and
varying $T$. This line passes through all three phases for thin
samples and is shifted to larger $\beta$ as $L_z$ is increased.}
\label{fig1}
\end{figure}

The phase diagram for Coulomb interacting bosons comprises three
phases and is shown schematically in Fig.~\ref{fig1}. The
parameter $\Lambda$ measures the size of quantum effects, and the
line $\Lambda = 0$ therefore corresponds to the classical 2D Coulomb
plasma. As the temperature is lowered, $\beta$ is increased and we
expect a transition from a fluid to a crystal. Numerical simulations
agree on this transition being first order, taking place at $\beta_m
\approx 140$%
\cite{Caillol-etal:JSP82,Leeuw-Perram:PhA82,Choquard-Clerouin:PRL83}. 
For large quantum effects, i.e., for large $\Lambda$, the low
temperature phase is a superfluid. The transition from a classical
fluid to a superfluid is generally accepted to be a
Kosterlitz-Thouless transition in two
dimensions\cite{Kosterlitz-Thouless:JPC73}. Finally, at large $\beta$,
the system goes from a lattice to a superfluid as the
size of quantum fluctuations is increased. Since no temperature is
involved, this is a quantum phase transition, and the critical value
of $\Lambda$ is $\Lambda_m \approx 0.062$ for $\beta =
\infty$\cite{Magro-Ceperley:PRL94}. This transition, which is
equivalent to the vortex lattice melting transition in a
superconductor, is the topic of the present study. 

The phase diagram can be conveniently discussed in terms of three
energy scales (potential, kinetic, and thermal) and two symmetries
(gauge and translational). As long as the thermal energy $T_B =
g^2/\beta$ dominates, the system possesses both global gauge
invariance and continuous translational symmetry. If the potential
energy $g^2$ dominates at low temperature, the translational symmetry
is broken and we find a crystal. Otherwise, if the quantum mechanical
kinetic energy $g^2\Lambda^2$ dominates, the gauge symmetry is broken
and the system becomes superfluid. Strictly speaking, since we are in
two dimensions, none of the symmetries is really broken at finite
temperatures. Rather, both the crystal and the superfluid phase are
characterized by \emph{quasi long-range} order, as will be discussed 
below.

According to the phase diagram in Fig.~\ref{fig1} two
symmetries must change simultaneously at the phase boundary between
the crystal and the superfluid. This has led to speculations on the
existence of other phases, such as the supersolid, in which both
symmetries are broken. The first suggestion for the existence of such 
a supersolid phase in the context of superfluid He II was made by
Andreev\cite{Andreev-Lifshitz:JETP69}, see
Ref.~\onlinecite{Meisel:PhB92} 
for an overview. The topic has also been studied in the context of
Josephson junction arrays, where a supersolid has been observed
numerically\cite{Otterlo-Wagenblast:PRL94}, and for vortices in
superconductors, where no supersolid was
found\cite{Frey-Nelson-Fisher:PRB94}. 

Another suggestion is due to the rather peculiar properties of the 2D
Bose Coulomb Liquid. It can be shown that the off-diagonal long-range
order in the ground state decays as 
\begin{equation}
\left\langle \psi^\dagger({\bf r}) \psi(0) \right\rangle \propto
r^{-\alpha}, 
\label{ODLRO}
\end{equation}
where the exponent $\alpha$ is given
by%
\cite{Feigelman-etal:PRB93,Magro-Ceperley:PRL94,%
Pitaevskii-Stringari:JLTP91}
\begin{equation}
\alpha = \frac 1 \Lambda \left( \frac{\sqrt{3}}{16\pi} \right)^{1/2}
= \left( \frac{mg^2}{8\pi\hbar^2 \rho} \right)^{1/2} \approx
\frac{0.186}{\Lambda}.
\label{ODLROexp}
\end{equation}
The system therefore does not have true off-diagonal long-range order
(ODLRO) even in the ground state. This is in contrast to the case
with short-range interactions where ODLRO decays algebraically only
at finite temperatures according to the Hohenberg-Mermin-Wagner
theorem\cite{Mermin-Wagner:PRL66}. The effect can be understood in
terms of the long-range interaction suppressing long wavelength
density fluctuations. Because of the uncertainty principle, this
implies that the fluctuations in the conjugate variable, the phase,
have to diverge. 

The question is how large $\alpha$ can become before superfluidity is
destroyed. One estimate is based on the analysis of the momentum
distribution function: Computing the Fourier transform of Eq.\
(\ref{ODLRO}) we obtain
\begin{equation}
n_{{\bf q}} \propto q^{\alpha - 2},
\end{equation}
and the momentum distribution function actually vanishes for
$q\rightarrow 0$ if $\alpha > 2$. Since the crystal melts at
$\alpha_m \approx 3$ ($\Lambda_m \approx 0.062$), it is relevant to
ask whether a 2D Bose liquid with $2 < \alpha < \alpha_m $ can be
superfluid. Otherwise, a non-superfluid Bose liquid would exist at
$T_B = 0$. One of the results of this work is to show that $\alpha$
can indeed be larger than 2 in a superfluid, and that superfluidity
only disappears when the lattice forms, confirming the conjecture
that a Bose system is either a superfluid or a crystal at $T_B = 0$.
Note that the concept of a superfluid without a Bose-Einstein
condensate is really not as striking as it might seem: A 2D Bose
system does not have a real condensate at finite temperatures but
can still be superfluid. Also, for He II in three
dimensions, the condensate fraction is below 10\% at zero
temperature, whereas the superfluid density equals the total
density\cite{Campbell:78}.

A further consequence of the long-range interaction is that the solid
phase appears at low densities, i.~e., it is a Wigner crystal. For a
finite interaction range $\lambda$, we also find a superfluid phase
at low densities with $\lambda < a_0$. Finally, we note that since
the interaction is purely repulsive no generic liquid exists; the
state which we refer to as a liquid might equally well be called a
gas or a plasma. 

It is instructive to discuss the phase diagram directly in terms of
world lines. The partition function (\ref{partfunc}) includes all
particle permutations, i.e., all ways of connecting the lower ends of
the world lines at $\tau = 0$ with the upper ends at $\tau = \beta$,
and most of the possible line configurations are heavily entangled.
There are two factors that prevent this entanglement: The length of
the world lines is determined by the temperature of the Bose system
and the world lines can therefore only entangle if the temperature is
low enough. A simple estimate for the entanglement temperature 
is given by
\begin{equation}
\frac{1}{\hbar}\frac{m a_0^2}{(\hbar/ T_B) } = \frac{1}{\Lambda^2
\beta} \approx 1,
\end{equation}
marking the appearance of superfluidity. Since the interaction
between the lines is also less effective at high temperatures, it
follows that the high-temperature phase is a normal liquid. If the
interaction is strong enough, it will prevent entanglement even of
infinitely long world lines, and we have a crystal ground state.

The above phase diagram can be reinterpreted as a phase diagram for
vortices by use of the parameter mapping in Eq.\ (\ref{parammap}).
The Wigner crystal corresponds to the vortex lattice, the superfluid
is equivalent to an \emph{entangled} vortex liquid, and the normal
fluid corresponds to a \emph{disentangled} vortex liquid. Since the
parameter $\beta$ is proportional to the thickness of the sample, we
are only interested in the limit $\beta \rightarrow \infty$ for a
bulk superconductor. It is important to use a large enough $L_z$ in a
numerical simulation: The dashed line in Fig.~\ref{fig1} shows
the curve obtained by keeping $B$ ($a_0$) constant and varying $T$
for some finite $L_z$. The curve passes from a lattice, via a
disentangled vortex liquid, into an entangled liquid. As $L_z$ is
increased, however, the curve is pushed to higher values of $\beta$
and we have a direct transition from a lattice to an entangled
liquid. 

\section{Notes on the algorithm}
\label{secalgo}

One of the advantages of using the Bose model for describing the
vortex system is that there exist well established algorithms for
simulating bosons numerically exactly. The ground state of a Bose
system can be studied using variational and diffusion Monte Carlo
methods, which has been done for the system considered here by
Ceperley and Magro\cite{Magro-Ceperley:PRB93,Magro-Ceperley:PRL94}.
They observed the disappearance of the Wigner lattice as $\Lambda$
was increased and also studied the momentum distribution function in
the liquid, which agreed with the result in Eqs.\ (\ref{ODLRO}) and 
(\ref{ODLROexp}). Simulations at $T_B = 0$ are usually not very good
at computing response functions, such as the superfluid density,
since these depend on the excitations available in the system rather
than on the ground state wavefunction. Although this difficulty can
be overcome for the special case of the superfluid
density\cite{Zhang-etal:PRL95}, we have chosen the more conventional
approach of simulating a system at finite temperature. Not only does
this allow us to compute the superfluid density using the winding
number, but it also allows us to study the excitation spectrum from
the dynamic structure factor.

Exact finite temperature simulations on Bose systems were introduced
by Ceperley and Pollock and applied to superfluid He II in two and
three dimensions%
\cite{Pollock-Ceperley:PRB84,Pollock-Ceperley:PRB87,%
Ceperley-Pollock:PRL89}. The method is known as the Path Integral
Monte Carlo (PIMC) method, and differs from the more widely used
world-line methods in that it does not introduce a lattice. It is
therefore possible to obtain numerically exact results for realistic
interactions between the bosons, as has been amply demonstrated by
Ceperley and co-workers in computing the superfluid density and the
excitation spectrum for superfluid He II\cite{Ceperley:RMP95}.

The method involves discretizing the imaginary time in the path
integral for the partition function, which, in the Trotter
approximation, becomes
\begin{equation}
{\cal Z}(\beta,\Lambda,N) = \frac 1 {N!} \left( \frac{1}
{2\pi\Lambda^2\tau} \right)^{MN}\!\int \prod_{m=1}^M
\prod_{i=1}^N d^2 R_{i,m} {\rm e}^{-S[\{{\bf R}_i]\}},
\label{discpart}
\end{equation}
with
\begin{equation}
S[\{{\bf R}_i\}] = \sum_{i,m} \frac{\left( {\bf R}_{i,m+1} - {\bf
R}_{i,m} \right)^2}{2\Lambda^2\tau} + \sum_{i<j,m} \tau {\rm K}_0
\left( R_{ij,m} /\lambda \right).
\label{discaction}
\end{equation}
Here the indices $(i,j)$ label the particles, $m$ labels the
Trotter slices ranging from $1$ to $M$, and we have $\tau = \beta /
M$ for the size of the imaginary time step. We have typically used $M
= 100$ and $N = 25-100$ in the simulations presented here, with
periodic boundary conditions in all directions. A rhombically shaped
system, with the smaller angle equal to $60^\circ$, has been used to
allow for a triangular lattice without frustration. Since the range
of the interaction is usually larger than the size of the system, it
is important to account for the periodicity of the system also in the
interaction. We do this by solving the London equation with the
correct boundary conditions and the resulting expressions are given
in Appendix~\ref{appfourier}.

Computation of thermodynamic averages using Eq.\ (\ref{discpart})
involves evaluating a multi-dimensional integral numerically. Since
the dimensionality is high, with $MN \approx 10^3 - 10^4$, an
efficient algorithm is needed. Most of the PIMC algorithm, including
the bisection method for generating new line configurations, is
described in the review by Ceperley\cite{Ceperley:RMP95}. There is
one point, though, where our algorithm differs from those used before:
In order to capture the quantum effects correctly in the Bose system,
it is necessary to sample the particle permutations accurately. Thus
we have to probe all ways of connecting the lower ends of the lines
${\bf R}_i(0)$ with the upper ends ${\bf R}_i(\beta)$. Usually this is
done by allowing a small number of lines to change their endpoints in
every Monte Carlo update. We were unable to equilibrate the system
using any such algorithm. Rather we use a random walk algorithm which 
allows the system to choose a permutation of a large number of lines 
according to its statistical weight. The algorithm is described in
Appendix \ref{apppermutation}. For N = 49 or less, we actually 
do this random walk in the space of the permutations of all $N$ lines, 
leading to a reasonably fast convergence of the winding number 
(see below).

\subsection{Thermodynamic properties}

We now turn to the measurable quantities of the system. The existence
of a lattice is tested by measuring the structure factor $S({\bf
Q})$,
\begin{eqnarray}
S({\bf Q}) & = & \frac 1 N \langle \rho({\bf Q}) \rho(-{\bf
Q})\rangle  
\nonumber \\ & = &
\frac{1}{MN} \left\langle \sum_{ij,m} \exp\left\{ i {\bf
Q} \cdot \left( {\bf R}_{i,m} - {\bf R}_{j,m} \right) \right\}
\right\rangle,
\label{strucfacdef}
\end{eqnarray}
where the same convention for the sums has been used as in Eq.\
(\ref{discpart}) and we use $\langle\dots\rangle$ to denote Monte
Carlo averages. The height of the structure factor evaluated at a
reciprocal lattice vector scales with the size of the system. More
specifically, we have
\begin{equation}
S\left({\bf Q}_1\right) = N\exp\left\{ -\frac{8\pi^2}{3a_0^2} \langle
u^2 \rangle \right\}
\label{debyewaller}
\end{equation}
for the first Bragg peak in the triangular lattice, where $\langle
u^2 \rangle$ denotes the mean-squared fluctuations. Expression
(\ref{debyewaller}) has the usual Debye-Waller form and requires the
lines to fluctuate harmonically around their average positions. 

We stated above that there should be no true long-range order in a
system of finite thickness ($\beta$). Consequently, there should not
exist real Bragg peaks. In order to understand this point better, we
compute $\langle u^2 \rangle$ using elastic theory for a system of
finite thickness. The main contribution comes from transverse
fluctuations, given by the expression
\begin{eqnarray}
\frac 1 2 \langle [u_\perp({\bf R}) &-& u_\perp(0)]^2\rangle 
\nonumber \\ 
& = &\frac{1}{L_z} \int \frac{d^2K}{(2\pi)^2} \sum_{k_{z}}\frac{T
\left[1 - \cos({\bf K}
\cdot{\bf R}) \right]}{c_{66}K^2 + c_{44}k_{z}^2}.
\end{eqnarray}
Here, the two-dimensional $K$ integral is over the Brillouin
zone, with $K_{BZ} = \sqrt{4\pi} / a_0$, and the discrete
$k_z$-vectors are given by $k_z = 2\pi n / L_z$. We ignore the
dispersion of the tilt modulus for the moment. Separating off the
$k_z = 0$ term and integrating over the remaining ones, we obtain
\begin{eqnarray}
\frac 1 2 \langle u_\perp({\bf R}) & - & u_\perp(0)]^2\rangle 
\nonumber \\ & \approx &
\frac{TK_{BZ}}{4\pi \sqrt{c_{66}c_{44}}}
+ \frac{T}{2\pi c_{66}L_z} \ln ( R / a_0 )
\label{fluctuations}
\end{eqnarray}
in the limit $R \gg a_0$. The first term is the result for the
transverse fluctuations in a bulk superconductor and contains
only $k_z \neq 0$ modes. The second term is due to the $k_z = 0$
mode and thus corresponds to motion of the average position of
a vortex line. In a sample of finite thickness there is always 
only quasi long-range order since the second term in 
Eq.~(\ref{fluctuations}) grows logarithmically with the 
distance $R$. This term, however, is only relevant if $R \gtrsim a_0
\exp\left( c_{66}a_0^2L_z /T\right)$, which grows exponentially with
the thickness of the sample. In the limit $L_z \rightarrow \infty$,
the second term of Eq.~(\ref{fluctuations}) vanishes and we recover
the well known result of true long-range order in the Abrikosov
lattice.

\begin{figure}
\centerline{\epsfxsize= 8.5cm\epsfbox{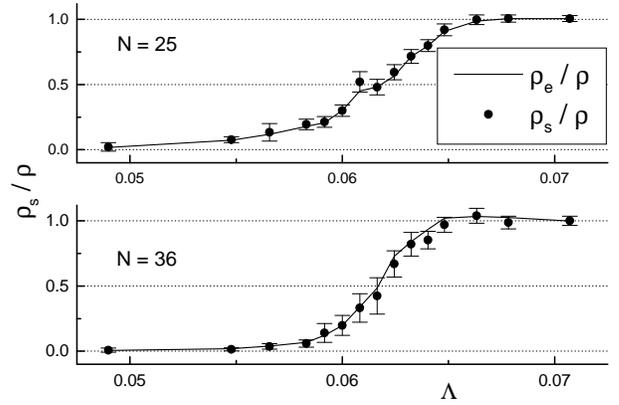}}
\caption{The superfluid density of the 2DBCL measured using the
winding number (dots) and the entanglement density (solid line) as
explained in the text. We have used the parameters $\beta = 300$, $M
= 100$, and $N = 25,36$. Both curves are fitted using the value
$\alpha(\Lambda,\beta) = 4.04$, which was established by requiring
the last point ($\Lambda \approx 0.071$) to have $\rho_e = \rho$.}
\label{fig2}
\end{figure}

We shall use the bosonic superfluid density as a measure of the
entanglement in the system. The superfluid density is defined
by the response of the system to a transverse gauge field, or,
equivalently, to the motion of the walls of the container. We
therefore have to study a system described by the Hamiltonian $H_{\bf
v}$, obtained from the original Hamiltonian by the transformation
${\bf p} \rightarrow {\bf p} - m{\bf v}$. The superfluid density is
then given by
\begin{eqnarray}
\frac{\rho_s}{\rho} & = & \frac{1}{mN} \frac{\partial^2}{\partial{\bf
v}\partial{\bf v}} F_{\bf v}(\beta,\Lambda) \nonumber \\
& = &  - \frac{T_B}{mN}
\frac{\partial^2}{\partial{\bf v}\partial{\bf v}} \ln {\cal Z}_{\bf
v}(\beta,\Lambda),
\label{sfdef}
\end{eqnarray}
where $F_{\bf v}$ and ${\cal Z}_{\bf v}$ represent the free energy
and the partition function for the system with moving walls. It can
be shown that the latter is given by\cite{Ceperley:RMP95}
\begin{equation}
{\cal Z}_{\bf v}(\beta,\Lambda) = {\cal Z}_{0}(\beta,\Lambda)
\exp\left\{ \frac{i m}{\hbar} {\bf W} \cdot {\bf v} \right\},
\label{Zv}
\end{equation}
where the winding vector $\bf W$ is defined as
\begin{equation}
{\bf W} = \sum_i \int d\tau \frac{d{\bf R}_i}{d\tau}.
\end{equation}
This definition deserves an explanation. Since the upper ends ($\tau
= \beta)$ are just a permutation of the lower ends ($\tau = 0$), the
only paths which give a contribution to the winding number are those
which leave the system on one side and return on another via the
periodic boundary conditions. For these paths we have 
\begin{equation}
\int d\tau \frac{d{\bf R}_i}{d\tau} = {\bf R}_i(\beta) - {\bf R}_i(0)
+ L{\bf n},
\end{equation}
where $L$ is the linear size of the system and ${\bf n} \in {\Bbb
Z}^D$ is an integer vector. We are only considering $D = 2$ in this
work, but the winding number definition of the superfluid density is
valid in any dimension $D$. Combining Eqs.~(\ref{sfdef}) and
(\ref{Zv}) we immediately arrive at
\begin{equation}
\frac{\rho_s}{\rho} = \frac{\langle W^2 \rangle}{D\Lambda^2\beta N},
\label{sfW}
\end{equation}
which is an exact relation. It turns out, however, that it is
numerically difficult to get good statistics for the winding number
in large systems since it can only be changed by global moves
involving
on the order of $\sqrt[D]{N}$ world lines. For large systems we then
made use of a simpler measure of quantum effects, which consists of
simply counting the number of world lines which do not end on
themselves:
\begin{equation}
N_e \equiv \text{number of lines with} \; {\bf R}_i(\beta) \neq {\bf
R}_i(0).
\end{equation}
This quantity does not need any global moves to be changed and is
therefore simple to measure for arbitrarily large systems. It is
interesting to consider the relation between $N_e$ and $\rho_s$.
Obviously $N_e > 0$ is a necessary condition for $\rho_s > 0$. There
is no guarantee, however, that the condition is also sufficient. In
order to arrive at a more quantitative relation between the two
quantities, we can reason as follows: The quantities $\rho_s / \rho$
and $N_e / N$ are intensive and should not depend on the total number
of particles in the system. If we assume that they are proportional
to each other, it follows from Eq.\ (\ref{sfW}) that
\begin{equation}
\langle W^2 \rangle \propto N_e,
\label{conjecture}
\end{equation}
which is reasonable, given that the vector ${\bf W}$ is
the sum of $N_e$ randomly directed vectors. Unfortunately, we cannot
offer a derivation of the proportionality constant in
Eq.~(\ref{conjecture}) and thus have to rely on
empirical results: We define the quantity $\rho_e$ according to 
\begin{equation}
\frac{\rho_e}{\rho} \equiv
\frac{\alpha(\Lambda,\beta)}{D\Lambda^2\beta} \frac{N_e}{N}
\label{sfN}
\end{equation}
and compute $\alpha(\Lambda,\beta)$ by requiring $\rho_e = \rho_s$
for small systems where $\rho_s$ can be computed using the standard
winding number definition (\ref{sfW}). It turns out that for the
range of parameters considered here, we can ignore the $\Lambda$ and
$\beta$ dependence of $\alpha(\Lambda,\beta)$ and use
\begin{equation}
\alpha(\Lambda,\beta) = 4.04.
\end{equation}
We emphasize, however, that $\alpha(\Lambda,\beta)$ does depend on
both $\beta$ and $\Lambda$ in general. Fig.\ \ref{fig2} shows
the superfluidity, measured according to Eqs.~(\ref{sfW}) and
(\ref{sfN}), as a function of $\Lambda$ for two different system
sizes. The good agreement between $\rho_e$ and $\rho_s$, particularly
the sharp onset of $\rho_e$ at melting (see Fig.\ \ref{fig4}), 
is a consequence of the strong
first-order character of the transition, and makes us confident that
$\rho_e$ provides a good estimate for the superfluid density in the
larger Bose systems.  Nonetheless, we stress that $\rho_e$ is not a
real order parameter since, in contrast to $\rho_s$, it does not 
vanish in the lattice phase when taking the thermodynamic limit.

The energy of the system, both for bosons and vortices, can be
computed using the thermodynamic definition
\begin{equation}
E = T^2 \frac{\partial}{\partial T} \ln {\cal Z}(\Lambda,\beta).
\label{Edef}
\end{equation}
Due to the different temperature dependence of the parameters
$\Lambda$ and $\beta$ in the two cases, the resulting expressions for
the energy are different. It is useful to define the two expectation
values
\begin{mathletters}
\begin{equation}
S_1 = \left\langle \sum_m \sum_i \frac{\left( {\bf R}_{i,m+1} - {\bf
R}_{i,m} \right)^2}{2\Lambda^2\tau}
 \right\rangle
\end{equation}
\begin{equation}
S_2 = \left\langle \sum_m \sum_{i<j} \tau {\rm K}_0 \left( R_{ij,m}
\right) /\lambda \right\rangle
\end{equation}
\end{mathletters}
corresponding to the two terms of the action (\ref{discaction}). The
important point is the temperature dependence of the term $S_1$. For
bosons, we have $\tau \propto 1 / T_B$ and $S_1 \propto T_B$. The
expression for the energy then becomes\cite{Ceperley:RMP95}
\begin{equation}
E_B = T_B( MN - S_1 ) + T_B S_2,
\label{BoseEnergy}
\end{equation}
where the two terms correspond to kinetic and potential energies,
respectively. The quantum-mechanical kinetic energy is decreased by
the fluctuations of the world lines, since the fluctuations
correspond to a smearing of the wave function. 

In the case of vortices, the denominator of $S_1$ is given by 
\begin{equation}
\Lambda^2 \tau \propto \frac{T}{\varepsilon_l}.
\end{equation}
If we want to treat ${\cal F}$ as a model Hamiltonian, we should
ignore its internal temperature dependence through the penetration
depth $\lambda(T)$. In this case, $\varepsilon_l$  and $\varepsilon_0$
are 
constants and both the terms $S_1$ and $S_2$ are
proportional to $1/T$. Computing the temperature derivative in
Eq.~(\ref{Edef}) we obtain  
\begin{equation}
\langle {\cal F} \rangle = T( S_1 - MN ) + T S_2,
\label{GLmean}
\end{equation}
where ${\cal F}$ is again the Ginzburg-Landau free energy functional.
The expectation value of ${\cal F}$ is a very useful quantity to
measure, but does not represent the energy of the vortex system. As
was pointed out by Hu and MacDonald\cite{Hu-MacDonald:PRB97}, the
internal temperature dependence of $\cal F$, through $\lambda(T)$,
 accounts for microscopic degrees of freedom that give a significant 
contribution to the energy. We will show below how Eq.~(\ref{GLmean})
can be combined with scaling properties of the London free energy to 
obtain the correct expression for the energy.

Another advantage of doing the simulation at finite temperature is
that it allows us to compute the excitation spectrum of the system.
The idea is to analyze the dynamic structure factor $S({\bf
Q},i\omega_n)$, which can be obtained as the partial Fourier
transform of the retarded density correlator 
\begin{equation}
S( {\bf Q}, \tau ) = \frac 1 N \left\langle \rho({\bf Q},\tau)
\rho(-{\bf
Q},0) \right\rangle.
\end{equation}
The normalization is chosen so that $S( {\bf Q}, \tau = 0 )$ is the 
static structure factor of Eq.~(\ref{strucfacdef}). 
The standard method for obtaining the excitation spectrum from
$S({\bf Q},i\omega_n)$ is to perform an analytic continuation to
real frequencies using the maximum entropy method (see
Ref.~\onlinecite{Jarrell-Gubernatis:PhysRep96} for an 
overview). Here we use a simpler approach, similar in spirit to
analytical work by Nelson and Seung\cite{Nelson-Seung:PRB89}. If we
assume the structure factor at momentum ${\bf Q}$ to be determined by
one quasi-particle excitation, we obtain the single-mode
approximation
\begin{equation}
S({\bf Q},i\omega_n) = \frac{C_{{\bf Q}}}{\left( w_n + \Gamma_{{\bf
Q}} \right)^2 + \varepsilon_{{\bf Q}}^2}.
\label{SingleMode}
\end{equation}
Here $\varepsilon_{{\bf Q}}$ and $\Gamma_{{\bf Q}}$ are the excitation
energy and inverse lifetime, respectively, of the quasi-particles,
and $C_{{\bf Q}}$ is related to the static structure factor of the
system. This form of the structure factor leads to an exponential
decay of correlations along the world lines according to $S( {\bf
{{\bf Q}}}, \tau ) \propto \exp( -\varepsilon_{{\bf Q}} |\tau| )$.

\begin{figure}
\centerline{\epsfxsize= 8.5cm\epsfbox{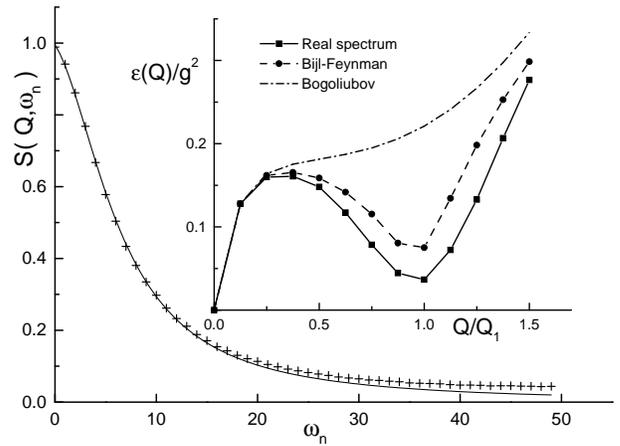}}
\caption{Fit of the single mode structure factor $S(Q,i\omega_n)$ to
measured data for a compressible system with $N=64$, $M = 100$, and
$\beta = 300$ at $Q/Q_1 = 0.5$. The fit is very good for low
frequencies and fails for higher ones as expected. The inset shows
the excitation spectrum obtained from our method compared to the
Bijl-Feynman and Bogolioubov spectra.}
\label{fig3}
\end{figure}

In order to understand the physical implications of the single-mode
approximation we consider the superfluid phase, where analytical
results are available. Eq.~(\ref{SingleMode}) is consistent with the
two standard sum rules, i.e., the fluctuation-dissipation theorem and
the $f$-sum rule\cite{Fetter-Walecka:71}, provided that 
\begin{mathletters}
\begin{eqnarray}
\varepsilon_{Q} & = & \frac{Q^2}{2m S(Q)}, 
\label{BF} \\
C_{Q} & = & 2  \varepsilon_{Q} S(Q),
\end{eqnarray}
\end{mathletters}
where $S(Q)$ is the static structure factor and all
quantities depend only on the magnitude of the wave vector ${\bf Q}$
since the superfluid is isotropic. The excitation spectrum
$\varepsilon_{Q}$ in Eq.~(\ref{BF}) is the Bijl-Feynman approximation,
which, according to the above derivation, is a direct consequence of
assuming one quasi-particle to satisfy the standard sum rules.
However, it is well known that the Bijl-Feynman approximation
significantly overestimates the energy of the excitations in 
the superfluid, 
and one excitation therefore does \emph{not} exhaust the sum rules.
Nonetheless, we can still assume the single-mode approximation to
give a valid description of the dynamic structure factor for small
frequencies, and use the form Eq.~(\ref{SingleMode}) to fit the
numerical data, as illustrated in Fig.~\ref{fig3}. The inset
shows the corresponding excitation spectrum, the Bijl-Feynman
spectrum computed from the static structure factor, and the
theoretical Bogoliubov result
\begin{equation}
\varepsilon^{B}_Q = \frac{\hbar^2}{2m}\sqrt{ Q^4 +
\frac{4m\rho}{\hbar^2}
Q^2 V(Q) }.
\end{equation}
The agreement is very good for small $Q$ but the Bijl-Feynman
approximation seriously overestimates the roton energy. We have not
performed an extensive analysis of the systematic errors in our
method and the results should therefore be considered as numerical
estimates rather than the final result on the excitation spectra in
2D Bose systems.

\subsection{Isobaric Monte Carlo}

A disadvantage of performing a simulation of a compressible
system at constant density is
that a first order transition will not be sharp. Rather, we will
observe the coexistence of the two phases at the transition,
complicating the interpretation of the data. As was shown in
Sec.~\ref{secmodel}, vortices in an applied magnetic field correspond
to bosons at fixed chemical potential, and a simulation using the
grand canonical ensemble would solve the problem with coexistence. It
is, however, technically difficult to vary the number of world lines
in the system. Furthermore, by varying the number of world lines, we
can only change the density in discrete steps, which is especially
problematic for small systems.

Since it is the density rather than the total number of particles
which is important, we can instead choose to simulate the Bose system
at constant applied pressure. The relevant thermodynamic potential is
then the Gibbs free energy, which is defined according to
\begin{equation}
{\cal Z}(T,P,N) \equiv {\rm e}^{-G(T,P,N)/T} = \int dA \, {\cal
Z}(T,A,N)\,{\rm e}^{-PA/T},
\label{partfuncP}
\end{equation}
in direct analogy to Eq.~(\ref{GrandCanDef}). The three thermodynamic
potentials $F(T,A,N)$, 
$\Omega(T,A,\mu)$, and $G(T,P,N)$ all offer perfectly valid
descriptions of the 2D Bose system, and the choice is therefore a
matter of convenience. Since $G(T,P,N)$ does not correspond directly
to the experimental situation of a superconductor in an applied
magnetic field, we have to make sure that the jump in density is
independent on whether we use $G(T,P,N)$ or $\Omega(T,A,\mu)$. We
therefore rewrite the free energy according to $F(T,A,N) =
Af(T,\rho)$, where $\rho = N/A$. The pressure and the chemical
potential are then fixed by the conditions
\begin{mathletters}
\label{TDdefs}
\begin{eqnarray}
P & = &\rho \frac{\partial f}{\partial \rho} - f, \\
\mu & = & \frac{\partial f}{\partial \rho}.
\end{eqnarray}
\end{mathletters}
We now consider two phases with free energy densities $f_1(T,\rho_1)$
and $f_2(T,\rho_2)$. The coexistence condition at fixed $\mu$
($\Omega$ continuous) is given by
\begin{equation}
\Delta f \equiv f_1 - f_2 = \mu \Delta \rho,
\end{equation}
where $\Delta \rho = \rho_1 - \rho_2$. The analogous condition at
fixed applied pressure ($G$ continuous) is given by
\begin{equation}
\Delta f = \frac{\Delta \rho}{\rho_1} \left( P + f_1 \right) =
\frac{\Delta \rho}{\rho_2} \left( P + f_2 \right).
\end{equation}
Comparison with Eq.~(\ref{TDdefs}) shows that in both cases we have 
\begin{equation}
\Delta f = \frac{\partial f}{\partial \rho} \Delta\rho,
\end{equation}
which is nothing but the Maxwell construction. The jump in density
will therefore be the same irrespective of whether we carry out the
simulation at constant $P$ or constant $\mu$. In the language of
vortices, this means that the magnetization jumps observed in the
simulation are directly comparable to experiments. It is, however,
not possible to deduce the strength of the applied magnetic field
from the pressure $P$, just as it is not possible to compute $\mu$
from $P$. According to Eqs.~(\ref{TDdefs}) we have $\rho\mu = P + f$,
which requires the knowledge of the free energy of the system.

The implementation of the isobaric quantum Monte Carlo algorithm is
similar in spirit to the classical case (see, e.~g., 
Ref.~\onlinecite{Allen-Tildesley:87}). Some details, which are mostly
of technical nature, are given in Appendix \ref{appisobaric}.

\section{Numerical results for the 2DBCL}
\label{sec2dbcl}

\begin{figure}[tb]
\centerline{\epsfxsize= 8.5cm\epsfbox{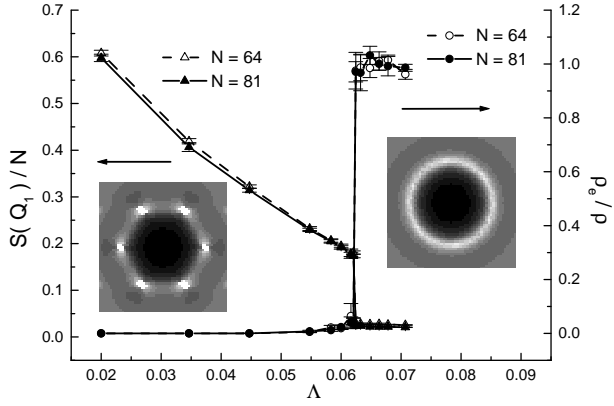}}
\caption{Structure factor and superfluid density at the transition
from a Wigner crystal to a superfluid for the 2DBCL. We have used
parameters $\beta = 300$, $M = 100$, and $N = 64,81$, and have
measured the superfluid density using $\rho_e$ as described in the
text. The peaks in the structure factor disappear in a sharp
transition at $\Lambda \approx 0.062$. Simultaneously, the superfluid
density rises from $\rho_e \approx 0$ to $\rho_e \approx \rho$. The
transition is very sharp, with a relative error $\Delta \Lambda /
\Lambda < 1\%$. In the language of vortices, the data shows a
first-order vortex lattice melting transition into an entangled
vortex liquid.} 
\label{fig4}
\end{figure}

\begin{figure}
\centerline{\epsfxsize= 8.5cm\epsfbox{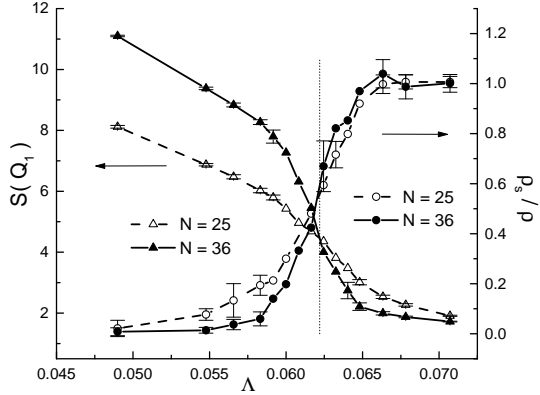}}
\caption{The same transition as in Fig.~\ref{fig4} for smaller
systems with $N = 25,36$. The superfluid density is measured directly
through the winding number. The vertical dotted line shows
the position of the transition for the larger systems in
Fig.~\ref{fig4}. The curves for the structure factor cross at
this line, showing that the position of the transition is not shifted
with the size of the system. The curves for the superfluid density
also cross very close to the vertical line, although the statistical
errors are larger. We have removed the error bars from some data
points for clarity.}
\label{fig5}
\end{figure}

One of the advantages of using the Bose model for studying the vortex
system numerically is that the algorithm can be tested against known
results for bosons. We begin by studying the transition from a Wigner
crystal to a superfluid in the 2D Bose Coulomb Liquid, i.e., with
$\lambda = \infty$. Fig.~\ref{fig4} shows the structure factor
and the superfluid density, computed according to Eq.~(\ref{sfN}), as
a function of $\Lambda$. We have used $\beta = 300$ and $M = 100$.
The lattice disappears at $\Lambda_m \approx 0.062$, in perfect
agreement with the results by Ceperley and Magro using a different
numerical technique for the same system\cite{Magro-Ceperley:PRL94}.
The transition is very sharp: The two structure factors shown as
insets correspond to $\Lambda = 0.06205$ and $\Lambda = 0.06245$, so
that the relative change in $\Lambda$ is less than 1\%.
Simultaneously, the superfluidity rises from $\rho_e = 0$ to $\rho_e
= \rho$, showing that the system is either superfluid or a crystal.
Since our method of measuring the superfluid density via $\rho_e$ is
not well established, we want to make sure that we obtain the
same results when measuring $\rho_s$ using the standard winding
number definition. Fig.~\ref{fig5} shows the same transition as
Fig.~\ref{fig4} for two smaller systems, where the winding
number has been measured directly. The dotted vertical line indicates
the position of the transition for the larger systems. The curves for
the structure factors cross exactly at this line, showing that the
position of the transition does not depend on the size of the system.
The curves for the superfluid densities cross very close to this line,
making us confident that there is only one transition in the system. 

\begin{figure}
\centerline{\epsfxsize= 8.5cm\epsfbox{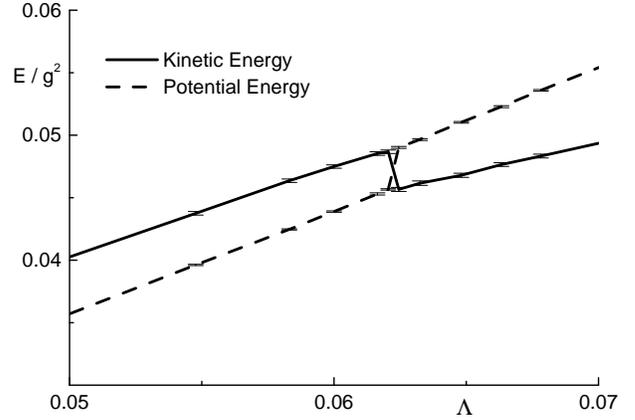}}
\caption{Potential and kinetic energies at the transition from
lattice to superfluid for the 2DBCL with $N = 81$ and $\beta = 300$.
The mismatch in energy is less than 10\% of the total jump, which is
comparable to the statistical error. The relative change in kinetic
energy at the transition is 7\%. The potential energy has been
shifted an arbitrary amount for clarity.}
\label{fig6}
\end{figure}

The transition from a Wigner crystal to a superfluid is a quantum
phase transition, which is driven by a change in $\hbar$ rather than
$T_B$. Since none of the two phases changes dramatically with
temperature, the properties of the transition will also not depend
strongly on temperature. This can be understood intuitively in the
world line picture: The world lines have to be long enough, but not
strictly infinitely long. There exists an elegant way to check
whether the temperature used in the simulation is low enough to
capture the ground state behavior: The free energy of the systems has
to be continuous across the transition, which implies
\begin{equation}
\Delta F = \Delta E_{kin} + \Delta E_{pot} - T_B \Delta S = 0.
\label{VeryClever}
\end{equation}
In particular, for the ground state we have $\Delta E_{kin} = -\Delta
E_{pot}$ since $T_B = 0$. Fig.~\ref{fig6} shows the kinetic
and potential energies for a system with $N = 81$. The transition
from a lattice to a superfluid is indicated by a sharp drop in the
kinetic energy, about $7\%$, due to the fact that the particles
become delocalized. This decrease in kinetic energy is compensated
by an increase in the potential energy. The almost perfect match
between kinetic and potential energies in Fig.~\ref{fig6} also
shows that the  systematic errors in the simulation due to the
discretization of the imaginary time are comparable to the
statistical error. Thus, we expect our results to hold for the
limit $\tau \rightarrow 0$, corresponding to continuous lines, and
for $\beta \rightarrow \infty$, corresponding to zero Bose
temperature. 

\begin{figure}
\centerline{\epsfxsize= 8.5cm\epsfbox{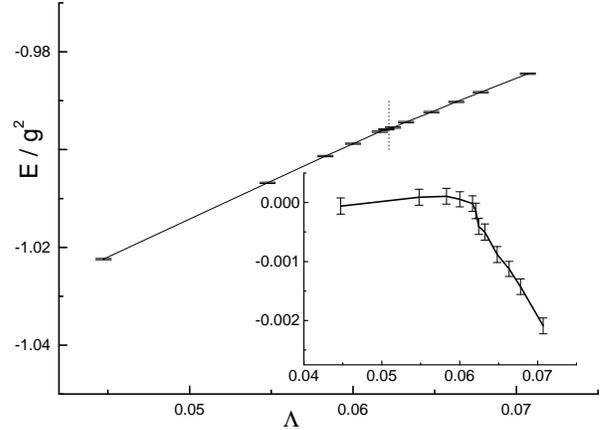}}
\caption{Energy of the 2DBCL as a function of $\Lambda$ for a system
with $N = 81$, $M = 100$, and $\beta = 300$. The transition,
indicated by the vertical dotted line, corresponds to a change in
slope. This can be seen more clearly in the inset, where we plot $E -
c\Lambda$ and have chosen $c$ to match the slope below the
transition.}
\label{fig7}
\end{figure}

In Fig.~\ref{fig7} we show the energy of the 2DBCL as a
function of the de Boer parameter $\Lambda$. The transition
corresponds to a small change in slope of this curve. The change is
more clearly seen in the inset, where we have subtracted the slope of
the crystalline phase. Strictly speaking, since we are simulating at
a finite temperature, there is a small downward jump in energy in
addition to the change of slope. The statistical errors are too large
to allow for a reliable estimate of this jump, however.

\section{Numerical results on vortex lattice melting}
\label{secmelting}

The above results on the transition in the 2DBCL can be directly
translated into results for vortex lattice melting. Thus,
Fig.~\ref{fig4} shows a first-order melting transition from a
lattice into an entangled vortex liquid. As long as we use the
approximation $\lambda = \infty$, the entire melting line is given by
the value of $\Lambda_m$, and we find
\begin{equation}
B_m(T) = \frac{\Phi_0}{\lambda^2} \frac{4 \Lambda_m^2}{\sqrt{3}}
\frac{\varepsilon_l\varepsilon_0\lambda^2}{T^2}.
\label{bosemeltline}
\end{equation}
This result is not applicable to the low field regime with $a_0 >
\lambda$, since the value of $\Lambda_m$ does depend on the range of
the interaction in this regime. For superconductors that are not too
anisotropic, the Lindemann criterion produces the melting
line\cite{Nordborg-etal:PRB96},
\begin{equation}
B_m(T) = \frac{\Phi_0}{\lambda^2} 4\pi c_L^2
\frac{\varepsilon^2\varepsilon_0^2\lambda^2}{T^2},
\label{Lindemannmeltline}
\end{equation}
which is identical to Eq.~(\ref{bosemeltline}) with $\varepsilon_l =
\varepsilon^2
\varepsilon_0 \ln( a_0 / 2\sqrt{\pi} \xi )$ and $\Lambda_m \propto
c_L^2$. We
therefore expect the Bose model to give a good description of vortex
lattice melting in YBCO, where the anisotropy, $\varepsilon \approx
1/7$, is moderate. For more anisotropic superconductors, such as
BSCCO, the electromagnetic coupling between pancake vortices becomes
important, an effect which is not accounted for by the Bose
model\cite{Glazman-Koshelev:PRB91,Nordborg-etal:PRB96}. It is
important to remember that the shape of the melting line is
mainly determined by the temperature dependence of the characteristic
line energy $\varepsilon_0$ rather than by details of the simulation.
We will therefore focus on the properties of the melting transition
itself, rather than on the shape of the melting line, in this work.

The structure factor shown in Fig.~\ref{fig4} is related to the
structure factor measured in small-angle neutron scattering (SANS)
experiments\cite{Cubitt-etal:NAT93}. The main difference is that our
structure factor measures the correlation in the position of the
vortex cores, whereas the neutrons are scattered by the magnetic
field of the vortices. This limits the resolution of SANS to length
scales larger than the magnetic penetration depth $\lambda$.

\begin{figure}
\centerline{\epsfxsize= 8.5cm\epsfbox{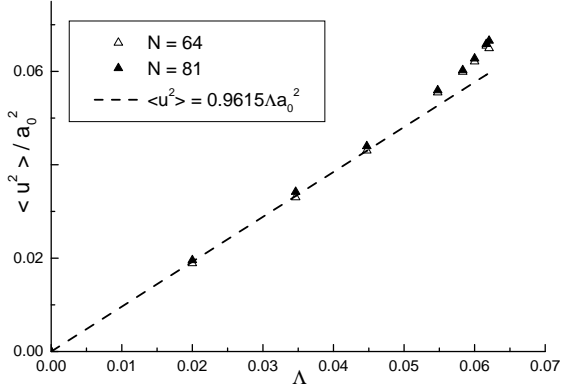}}
\caption{Mean squared fluctuations of in the vortex system. The data
points were obtained from the structure factor in
Fig~\ref{fig4}. The dashed line is a fit to the data for small
$\Lambda$ and gives $\langle u^2 \rangle \approx 0.9615 \, \Lambda
a_0^2$. The simple linear behavior breaks down for $\Lambda > 0.05$
due to renormalization of the elastic moduli. The Lindemann number
computed from the size of the fluctuations just before melting is
$c_L \approx 0.25$.}
\label{fig8}
\end{figure}

The elastic theory for the Bose model of the vortex lattice, i.e.,
with the elastic moduli given in Sec.~\ref{secvalidity}, predicts
$\langle u^2 \rangle \propto \Lambda$. Using Eq.~(\ref{debyewaller})
it is possible to measure the mean-squared fluctuations numerically,
and the result is shown in Fig.~\ref{fig8}. The dashed line is
a fit to the data for small $\Lambda$ and we find
\begin{equation}
\langle u^2 \rangle \approx 0.9615 \, \Lambda a_0^2.
\end{equation}
As is seen from the fit, elastic theory works well for $\Lambda
\lesssim 0.05$. Closer to the melting transition the renormalization
of the elastic moduli becomes important, and the linear dependence of
the fluctuations on $\Lambda$ breaks down. Just below the melting
transition we find $\langle u^2 \rangle \approx 0.065$, leading to a
Lindemann number of
\begin{equation}
c_L \approx 0.25.
\end{equation}
This result is in agreement with usual estimates for vortex lattice
melting\cite{Blatter-etal:RMP94}. 

The amount of entanglement is always very small in the lattice: We
typically find  $N_e / N \approx 0.026$ just below the transition,
compared to $N_e / N \approx 0.56$ in the liquid just above the
transition. Entanglement therefore increases by a factor 20 at the
transition already for a system with $N = 81$, and we have direct
melting into an \emph{entangled} vortex liquid. This is in agreement
with recent numerical results on other
models\cite{Hu-etal:PRL97,Koshelev:PRB97}, and with flux transformer
experiments\cite{Lopez-etal:PRL96}. 

In order to further characterize the melting transition we consider
the energy, which was defined in Eq.~(\ref{Edef}). For the case of
vortices, where ${\cal Z} = \exp\left( - {\cal F}/T \right)$, we
obtain
\begin{equation}
E = \langle {\cal F} \rangle - T \left\langle \frac{\partial {\cal
F}}{\partial T} \right\rangle.
\label{Evortex}
\end{equation}
The second term in this expression accounts for the internal
temperature dependence in the effective free energy and was neglected
in early numerical work on vortex lattice melting, which lead to much
too small estimates for the latent heat of the transition.

\begin{figure}
\centerline{\epsfxsize= 8.5cm\epsfbox{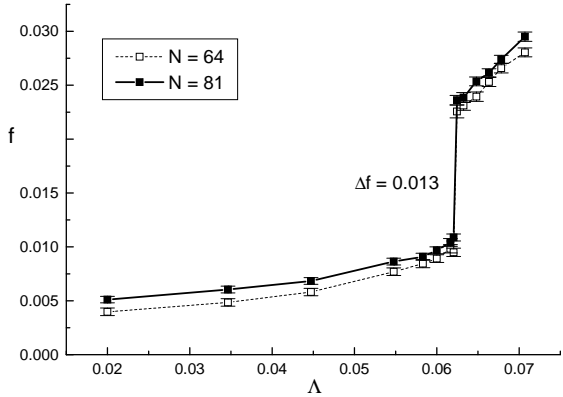}}
\caption{The GL free energy per particle and length for vortices with
$\lambda = \infty$. The energy of a perfect vortex lattice has been
subtracted. The melting transition is clearly seen as a discontinuity
in the energy, $\Delta f \approx 0.013$. The same systems as in
Fig.~\ref{fig4} were used.}
\label{fig9}
\end{figure}

Although it would be possible to measure the quantity in
Eq.~(\ref{Evortex}) directly in a simulation, we adopt a simpler
approach here\cite{Dodgson-etal:PRL98}. By direct inspection of the
free energy (\ref{GLfree}) we see that we can make use of the scaling
form
\begin{equation}
{\cal F}(\{ R_i(\tau)\},T) \approx \varepsilon_0(T) a_0 \,
\tilde{\cal F}(\{ R_i(\tau)\}),
\label{scaling}
\end{equation}
where $\tilde{\cal F}$ is dimensionless and temperature independent.
The approximation only neglects the temperature dependence of 
$\lambda(T)$ in the interaction, and Eq.\ (\ref{scaling}) therefore
becomes exact in the limit $\lambda \rightarrow \infty$. Combining
Eqs.\ (\ref{Evortex}) and (\ref{scaling}) we obtain the result
\begin{equation}
E = \langle {\cal F} \rangle \left( 1 - \frac{T}{\varepsilon_0}
\frac{d\varepsilon_0}{dT} \right) = \langle {\cal F} \rangle\frac{1 +
t^2}{1 - t^2},
\label{energyform}
\end{equation}
where we have assumed $\varepsilon_0(T) = \varepsilon_0(0) (1 - t^2)$
for the last equation. In the simulations, we always measure the
quantity $\langle {\cal F} \rangle$ and express all results in terms
of 
\begin{equation}
f  \equiv \frac{\langle {\cal F} - {\cal F}_0 \rangle}{N L_z
\varepsilon_0(T)},
\label{fdef}
\end{equation}
where ${\cal F}_0$ is the GL free energy for a perfect lattice of
lines. Thus, $f$ measures the additional GL free energy, due to
thermal fluctuations, per vortex and length in units of
$\varepsilon_0(T)$. The real energy of the system can then be easily
computed using Eq.\ (\ref{energyform}). 
In Fig.~\ref{fig9} we show the GL free energy of the vortex system
as a function of $\Lambda$. The transition is clearly seen as a
discontinuity in $f$ with $\Delta f \approx 0.013$. Using
Eq.~(\ref{energyform}), we compute the entropy per vortex and layer
and obtain  \begin{equation}
\Delta S = 0.013\, (1+t^2) \frac{\varepsilon_0(0) d}{T_m},
\end{equation}
where $d$ is the layer separation. With parameters for YBCO ($d
\approx 12 {\rm \AA}$, $\lambda_{ab} \approx 1400 {\rm \AA}$, and
$T_m\approx T_c \approx 92 K$) we obtain $\Delta S \approx 0.35 k_B$,
in rough agreement with experimental
results\cite{Schilling-etal:NAT96}.

It is interesting to consider how much of the latent heat comes from
the Josephson energy in our simulation, i.e., from the first term in
Eq.~(\ref{GLfree}). It has been shown numerically that approximately
half of the latent heat comes from the Josephson energy in a model
for a layered superconductor, a relation which is expected to become
exact for a line model\cite{Koshelev:PRB97}. Here we reach the same
conclusion in a slightly different way: Comparing Eqs.\
(\ref{BoseEnergy}) and (\ref{GLmean}) we see that the bosonic
kinetic energy is the negative of the Josephson energy. Since it was
shown above that the changes in kinetic and potential energies for
the quantum system have to cancel exactly, it follows that the change
in the Josephson energy equals the change in the in-plane interaction
energy. 

There exist a number of methods for testing numerically whether a
transition is first order. A widely used method is to make a
histogram of the data for the energy and look for two peaks,
indicating the coexistence of two phases\cite{Lee-Kosterlitz:PRL90}.
The method did not work very well in the present case, since the
transition is too sharp: Coexistence of phases was only observed for
small systems, where it is difficult to resolve the two peaks in the
energy distribution. The sharpness of the transition for large system
is sufficient evidence that the transition is really first order. 

\begin{figure}
\centerline{\epsfxsize= 8.5cm\epsfbox{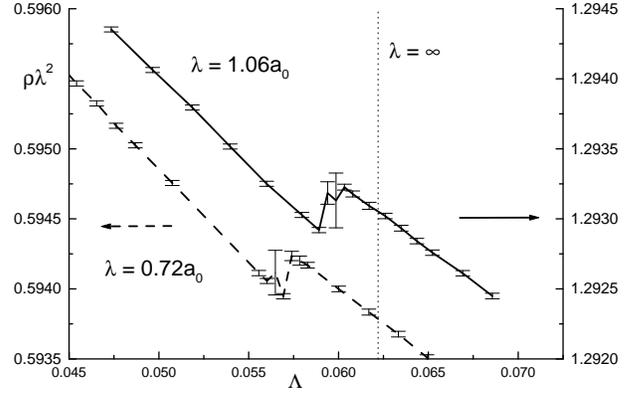}}
\caption{The change in density for two systems simulated with the
isobaric Monte Carlo method. We have used the parameters $\beta =
300$, $M = 100$, and $N = 64$. The vertical dotted line indicates the
position of the transition in the 2DBCL. The transition is shifted to
smaller values of $\Lambda$ as a result of the weaker interaction.}
\label{fig10}
\end{figure}

We now turn to the case of finite $\lambda$, where the system has a
finite compressibility. According to the Clausius-Clapeyron relation, 
\begin{equation}
\Delta s = - \frac{\Delta B}{4\pi} \frac{dH_m}{dT},
\label{ClausClap}
\end{equation}
we now expect a change in the density of the vortex system at the
transition. When going from the lattice to the liquid phase we have 
$dH_m/ dT < 0$ and $\Delta s > 0$, leading to $\Delta B > 0$. The
vortex system therefore shows anomalous melting, just as ice, with
the melted phase being the denser one. In the case of the vortex
system, the reason for the anomalous behavior can be understood in an
intuitive way: When the lines begin to entangle, the line tension
will produce an attractive force between them, leading to an
increased density\cite{Nelson:NAT95}. If we assume $H_m \approx B_m$,
which is true as long as $\lambda \gtrsim a_0$, we can rewrite
Eq.~(\ref{ClausClap}) as
\begin{equation}
\Delta f = 8\pi \Delta \rho \lambda(T)^2.
\label{SimClausClap}
\end{equation}
Fig.~\ref{fig10} shows the jump in density for two different
values of $\lambda$. The interaction between the lines is 
smaller than for $\lambda = \infty$ and the melting transition is
shifted towards lower values of $\Lambda_m$. 
The results are summarized in Table~\ref{tab1}. 
There is no dramatic change in the properties of the transition when
going from $\lambda = \infty$ to $\lambda \approx a_0$. This shows
that the logarithmic vortex interaction is a good approximation for
most of the phase diagram. The agreement with the Clausius-Clapeyron
relation in Eq.~(\ref{SimClausClap}) is very good for $\lambda = 1.06
a_0$, proving that the approximation $H_m \approx B_m$ is valid and
that the  transition is really first order. For $\lambda = 0.72a_0$,
the right hand side of Eq.~(\ref{SimClausClap}) is too small, since
$|dB/dT| < |dH/dT|$. Using the fact that $\Delta f$ does not depend
strongly on $\lambda / a_0$, we can use the value $\Delta f \approx
0.013$ and obtain a very simple expression for the density change of
the vortex lattice at the melting transition,
\begin{equation}
\Delta \rho \approx 5.2\times 10^{-4} / \lambda(T)^2
\Rightarrow\Delta B = \frac{1.1\times 10^{6} \,[{\rm G}{\rm
\AA^2}]}{\lambda(T)^2}
\end{equation}
This expression is again in good agreement with magnetization
measurements on YBCO\cite{Welp-etal:PRL96}.

\section{Properties of the line liquid}
\label{secliquid}

We now turn to the properties of the vortex liquid. It has been shown
above that the vortex lattice melts directly into an entangled vortex
liquid, which is equivalent to a bosonic superfluid. One important
quantity in the vortex liquid is the correlation length $l_z$ in the
direction parallel to the lines. If this correlation length grows
very large close to the melting line, an apparent disentangled
liquid could be observed in a sample of finite thickness. On the
other hand, we want the correlation length to be larger than the 
layer distance in order to separate melting of a line system 
from decoupling. A lot
of experimental effort has been invested into measuring $l_z$,
mainly using flux transformer techniques. The latest results indicate
that there exists a short but finite $z$-axis correlation length in
the vortex liquid\cite{Righi-etal:PRB97}.

One of the nice features of the Bose model is that the question
of the $z$-axis correlation length in the vortex system is related to
the fundamental question of the roton minimum in a Bose system. In
Sec.~\ref{secmodel} we have shown that the coordinate $z$ along the
vortex lines corresponds to the imaginary time $\tau$ for bosons.
This relation can be made dimensionally correct by writing
\begin{equation}
\frac{g^2 \tau}{\hbar} = \frac{2\varepsilon_0 l_z}{T}.
\label{timelength}
\end{equation}
According to Eq.~\ref{SingleMode} we expect the retarded density
correlator to decay exponentially, with the characteristic time
$\tau_Q = \hbar / \varepsilon_Q$. Using Eq.~(\ref{timelength}) we
obtain\cite{Nelson-Seung:PRB89}
\begin{equation}
l_z(Q) = \frac{T}{2\varepsilon_0} \frac{g^2}{\varepsilon_Q}.
\end{equation}
The largest correlation length is found for the $Q$ that minimizes
$\varepsilon_Q$, i.e., for the roton minimum. Since the roton minimum
is
located at $Q \approx 2\pi/a_0$, this length is related to the single
line correlation length. 

\begin{figure}
\centerline{\epsfxsize= 8.5cm\epsfbox{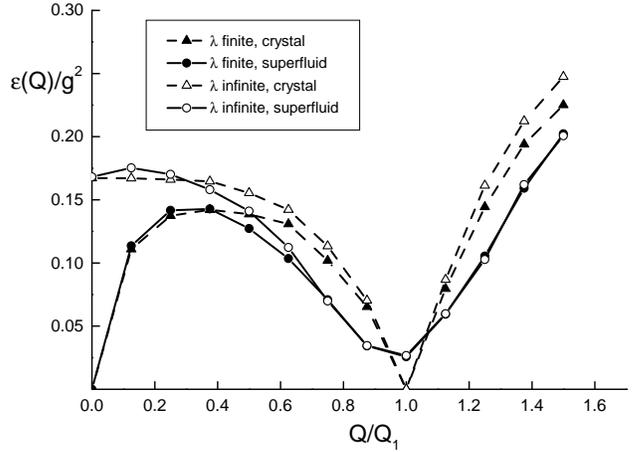}}
\caption{Comparison of the excitation spectra for compressible and
incompressible systems. The parameters are $\beta = 300$, $M = 100$,
$N = 64$, and the compressible system has $\lambda \approx 1.06a_0$.
The compressible system has a phonon branch for small $Q$. As the
range of interaction is increased, the phonons turn into plasmons.
The roton minimum is independent of the range of interaction, but
collapses as the system crystallizes.}
\label{fig11}
\end{figure}

Fig.~\ref{fig11} shows the excitation spectra for compressible
and incompressible 2D Bose systems. At small $Q$, the compressible
system has a phonon branch, whereas the excitations in the
incompressible systems are plasmons. The roton minimum, however, does
not change with the range of interaction. Since it is the roton
minimum which determines the critical velocity of the superfluid,
this supports our claim that the 2DBCL does not differ qualitatively
from other 2D Bose liquids with regard to superfluidity. The roton
minimum collapses when the system solidifies, indicating 1) the
appearance of a periodic lattice and 2) the disappearance of
superfluidity. Note that the roton energy is large compared to the
Bose temperature used in the simulation, $T_B \approx 0.003\, g^2$
compared to $\varepsilon_{roton} \approx 0.026\, g^2$. This explains
why we
observe the ground state behavior in the simulation. 

Recalculating the roton minimum in terms of Eq.~(\ref{timelength}) we
find 
\begin{equation}
l_z\left( \frac{4\pi}{a_0 \sqrt{3}} \right) \approx \frac 1 {0.026}
\frac{T_m}{2\varepsilon_0} \approx 1.7 \sqrt{
\frac{\varepsilon_l}{\varepsilon_0}}a_0
\approx 1.7 \,\varepsilon a_0.
\end{equation}
This result is in good agreement with our expectations from elastic
theory and scaling theory: The relevant length scale in the lattice
is $a_0$, which should be rescaled by anisotropy. The length $l_z$ is
much larger than the layer spacing for YBCO, proving that the
transition is really due to melting of continuous lines. As soon as
the lattice is formed, $l_z$ becomes infinite and entanglement
disappears.

\section{Conclusions}
\label{secconclude}

We have used the mapping to 2D bosons for studying the vortex system
in a high-$T_c$ superconductor numerically. An advantage of this
approach is that it allows us to perform the simulation without
introducing an artificial lattice, which would cut off small scale
fluctuations and introduce spurious pinning effects. Furthermore, by
comparing the numerical results to known properties of the
corresponding Bose system, we are able to control the systematic
errors inherent in any numerical simulation. 

We find a first-order vortex lattice melting transition into an
entangled vortex liquid. The results for the latent heat and the
magnetization jump agree well with experiments for moderately
anisotropic superconductors such as YBCO. We do not expect such good
agreement for strongly anisotropic materials, since the Bose model
describes vortices in terms of elastic lines, rather than pancake
vortices, and ignores the electromagnetic coupling between the
layers. 

In the language of bosons, we have found a sharp transition from a
Wigner crystal to a superfluid even for the case of a logarithmic
interaction (the 2DBCL). Since the latter system does not have a
Bose-Einstein condensate even in the ground state, this shows that
the relation between superfluidity and Bose-Einstein condensation is
rather subtle. It is amusing to note that the main result in this
work can be predicted using two simple physical ideas: 1) the Landau
quasi-particle picture and 2) the Bijl-Feynman form for the
excitation spectrum in a Bose system. 

The emphasis of this work has been on the generic phase diagram for a
system of interacting elastic lines. In this picture, the transition
described above can be understood in terms of a competition between
``energy'' and ``entropy'': The ordered lattice always has the lowest
``energy'', but there are many more entangled states. The
entanglement also offers a natural explanation for the anomalous
melting characteristic of vortex lattice. In order for this scenario
to work, however, the lines have to be long enough. Thus, one of the
main results of this work is to elucidate the importance of the third
dimension for vortex lattice melting.

\section*{Acknowledgments}

The authors are pleased to acknowledge fruitful discussions with
M.~Dodgson, M.~Feigel'man, E.~M. Forgan, V.~Geshkenbein, L.~Ioffe,
A.~Koshelev, A.~van Otterlo, H.~Tsunetsugu, and V.~Vinokur. The
simulations were carried out using the workstation cluster at the
Institute for Theoretical Physics, ETH, and on various computers of
the Centro Svizzero di Calcolo Scientifico. Support from the Swiss
National Science Foundation is gratefully acknowledged. Work at ANL
was funded in part by the NSF-Office of Science and Technology
Centers
under contract No.~DMR91-20000.

\appendix

\section{Computing the Fourier sum for the interaction}
\label{appfourier}

We consider a rhombically shaped system with side $L$ and the smaller
angle $\theta$. In the simulation we have used $\theta = 60^\circ$,
but nothing is lost by solving the more general problem here. Since
we want to be able to choose the range of the interaction $\lambda$
large compared to the size of the system, it is important to account
for the periodic boundary conditions in the interaction. In the
special case of the vortex interaction this is most easily achieved
by solving the London equation using the correct periodic boundary
conditions. We introduce a basis ${\bf e}_1, {\bf e}_2$ with ${\bf
e}_1 \cdot {\bf e}_2 = \cos \theta$. The system is then spanned by
the vectors ${\bf a}_i = L {\bf e}_i$, and the corresponding
reciprocal lattice
vectors are
\begin{equation}
{\bf b}_i = \frac{2\pi}{L \sin^2 \theta} \left( {\bf e}_i - {\bf e}_j
\cos \theta \right)
\end{equation}
for $(i,j) = (1,2),(2,1)$. The solution to the London equation
\begin{equation}
( 1 - \lambda^2 \Delta) G({\bf R},\lambda) = \lambda^2\delta({\bf R})
\end{equation}
is then given by
\begin{equation}
G({\bf R},\lambda) = \frac{2\pi\lambda^2}{L^2 \sin \theta} \sum_{{\bf
G}} \frac{{\rm e}^{i{\bf G}\cdot {\bf R}}}{1 + \lambda^2 G^2},
\label{FourSum}
\end{equation}
where the sum is over all reciprocal lattice vectors ${\bf G} =
n_1{\bf b}_1 + n_2{\bf b}_2$. This solution is proportional to the
magnetic field from one vortex line and in the case of $L \gg
\lambda$, the function $G({\bf R},\lambda)$ reduces to ${\rm K}_0(
R/\lambda)$. Since there is no such simplification for general $L$,
we will have to evaluate the Fourier sum in Eq.~(\ref{FourSum})
numerically. It is, however, very time consuming to sum over $n_1$
and $n_2$ directly, and we have therefore decided to evaluate one of
the sums analytically. This can be done with some effort using
contour integration, and the resulting expression is
\begin{equation}
G({\bf R},\lambda) = \frac{\sin\theta}{2(1 + \cos\theta)} \sum_{N}
\cos(N t_1) C_N(\lambda,t_2),
\label{fastsum}
\end{equation}
with $t_{1,2} = \pi ( R_1 \pm R_2 )/ L$ and
\begin{equation}
C_N(\lambda,t_2 ) =
\renewcommand{\arraystretch}{2.5}
\left\{
\begin{array}{cc}
\displaystyle \frac{\cosh[(\pi/2-|t_2|)a_N(\lambda)]}
{a_N(\lambda) \sinh[\pi a_N(\lambda) /2]}, &  N \quad\text{even}, \\
\displaystyle \frac{\sinh[(\pi/2-|t_2|)a_N(\lambda)]}
{a_N(\lambda) \cosh[\pi a_N(\lambda) /2]},  &  N \quad\text{odd},
\end{array}
\right.
\end{equation}
where
\begin{equation}
a_N(\lambda)^2 = \frac{L^2}{2\pi^2\lambda^2} \frac{\sin^2\theta}{1 +
\cos\theta} + \frac{1 - \cos\theta}{1 + \cos\theta} N^2.
\end{equation}
Eq.~(\ref{fastsum}) replaces an algebraically converging
two-dimensional sum with an exponentially converging one-dimensional
one, making the numerical evaluation trivial.

\section{Sampling permutations}
\label{apppermutation}

In order to capture the effects of Bose statistics, we have to allow
the world lines to switch their endpoints. This can only be done by
cutting out sufficiently long segments of a number of lines and 
trying different ways of connecting the loose ends. Updating the lines
therefore becomes a two-step process: First we decide on how to
connect the lower ends of the cut with the upper ones and then the 
new line segments are built up using the bisection method proposed by
Ceperley\cite{Ceperley:RMP95}. 

The problem of connecting the upper and lower ends of a cut of $N$ 
world lines is difficult for the simple reason that there 
exist $N!$ ways of doing it.
One way around the problem would be to choose only a small number of
neighboring lines for the Monte Carlo updates, thus keeping $N$ small.
On the other hand, we know that superfluidity is due to large scale
particle exchanges; the number of particles participating in a
Monte Carlo update has to be large enough to change the winding number
of the system. Different approaches have been suggested for solving
this problem and some of them can be found in
Ref.~\onlinecite{Ceperley:RMP95}. 

The method we use for connecting the ends of the cut is a random
walk in the space of permutations. The statistical weight of
interconnecting lines is only determined by the kinetic part of the
action and we introduce the weight
\begin{equation}
w_{ij} = \exp\left\{ -\frac{\left[ {\bf R}_j(\tau + \Delta) - {\bf
R}_i(\tau)\right]^2}{2\Lambda^2\Delta} \right\}
\end{equation}
for connecting the lower end of line $i$ with the upper
end of line $j$ over a distance $\Delta$ in the imaginary time
direction. The total weight of a permutation $\sigma$ of $N$ lines is
then given by
\begin{equation}
W(\sigma) = \prod_{i = 1}^{N} w_{i\sigma(i)}.
\end{equation}
The probability of the same permutation is the weight normalized by
the sum of the weights of all possible permutations, and this
normalization factor is practically impossible to compute if $N$ is
large. The random walk algorithm selects a permutation in the 
following way: 
We first select how to connect the first line with a probability
proportional to the weights. Thus, the probability of connecting
the lower end of the first line with the upper end of line $j$ is
\begin{equation}
p_{1j} = \frac{w_{1j}}{\sum_{k = 1}^N w_{1k}}.
\end{equation}
After this we select how to connect the second line to any of the
remaining $N-1$ lines, again with the probability proportional to the
weight. The procedure is repeated until all lines have been connected.
The acceptance probability for this permutation is then given by
\begin{equation}
A(\sigma) = \prod_{k = 1}^N \frac{\sum_{l = k}^{N}
w_{k\sigma(l)}}{\sum_{l = k}^{N} w_{kl}}.
\end{equation}
If the permutation is not accepted, the random walk is started again.
The algorithm is exact as long as the identity permutation is the
cheapest permutation, i.e., as long as $A(\sigma) \le 1$ for all
possible permutations, something which is almost always true even in
the superfluid phase. It is still not possible to sample the
permutations of arbitrarily large systems since the acceptance rate
decreases with $N$ and it becomes necessary to try very many random
walks before an acceptable permutation is found. However, since the
weights $w_{ij}$ can be computed once and stored in a table, each
random walk is very fast. Using this algorithm we were able to sample
the permutations of $N \le 49$ particle almost exactly. For larger 
systems, we typically restricted ourselves to $N \approx 25$ 
neighboring lines.

\section{Isobaric Monte Carlo algorithm}
\label{appisobaric}

In order to implement the isobaric Monte Carlo method, we first
rewrite the discretized action (\ref{discaction}) using the system
size $L$ as a length scale,
\begin{eqnarray}
S[\{{\bf x}_i\},L] & = & L^2\sum_{i,m} \frac{\left( {\bf x}_{i,m+1} -
{\bf x}_{i,m} \right)^2}{2\Lambda^2\tau} - MN\ln(L^2)\nonumber \\
& + & \sum_{i<j,m} \tau {\rm K}_0 \left( \frac{Lx_{ij,m}}{\lambda}
\right) + \beta P L^2.
\label{isobaricaction}
\end{eqnarray}
Here we have added a term corresponding to the external pressure
(cf.\ Eq.~(\ref{partfuncP})) and a logarithmic term coming from the
change of variable, ${\bf R} = L{\bf x}$, in the integral 
\begin{equation}
\int \prod_{m=1}^M \prod_{i = i}^{N} d^2 R_{i,m} = {\rm
e}^{MN\ln(L^2)} \int \prod_{m=1}^M \prod_{i = i}^{N} d^2 x_{i,m}.
\end{equation}
Using Eq.~(\ref{isobaricaction}) we can compute the change in all
thermodynamic quantities as a function of the area of the system. The
only problem is with the interaction term, which does not scale in a
simple way with the area. In the present simulation, the interaction
is stored in a lookup table and it would be too time consuming to
reinitialize these tables every time the volume changes. We are
helped by the fact that the volume changes are small, allowing us to
use a Taylor expansion,
\begin{equation}
U\left[L^2(1 + \varepsilon) \right] \approx U(L^2) + L^2\frac{d
U}{d(L^2)}\varepsilon +\frac 1 2 L^4\frac{d^2
U}{d(L^2)^2}\varepsilon^2,
\end{equation}
where
\begin{equation}
U(L^2) = \sum_{i<j,m} \tau {\rm K}_0 \left( Lx_{ij,m}/\lambda
\right).
\end{equation}
The interaction and its two derivatives with respect to the area are
evaluated for fixed $L$ during the simulation.  As long as the
changes in $L$ are small, we can use the Taylor expansion for
evaluating the change in Eq.~(\ref{isobaricaction}).

Rather than sampling the fluctuations in $L$, we used 
Eq.~(\ref{isobaricaction}) to optimize the action with respect to $L$
after the line configuration has been updated once. This gives the
correct expectation value for the density and the other thermodynamic
quantities.


\begin{table}
\caption{Jump in energy and density for three systems with $N = 64$, 
$\beta = 300$, and $M = 100$ at different pressures. The relative 
error is less than $5\%$ in the measured quantities }
\label{tab1}
\begin{tabular}{ccccc}
$P$ &  $\lambda / a_0$ & $\Lambda_m$ & $\Delta f$ & $8\pi \lambda^2
\Delta
\rho$ \\
\hline
$\infty$ & $\infty$ & 0.0622 & 0.0127 & -- \\
5 & 1.06 & 0.0596 & 0.0113 & 0.0116  \\
1 & 0.72 & 0.0567 & 0.0091 & 0.0085 \\
\end{tabular}
\end{table}

\end{document}